\documentclass[12pt]{article}

\usepackage[dvips]{graphicx}
\usepackage{epsfig}
\usepackage{amsmath}
\usepackage{amssymb}
\usepackage{float}

\newcommand{\bm}{\begin{multiline}}
\newcommand{\beq}{\begin{equation}}
\newcommand{\eeq}{\end{equation}}
\newcommand{\beqs}{\begin{eqnarray}}
\newcommand{\eeqs}{\end{eqnarray}}

\newcommand{\ra}{\rightarrow}

\begin{document}

\thispagestyle{empty}

\hfill{}

\hfill{}

\hfill{}

\vspace{32pt}

\begin{center}

\textbf{\Large On squashed black holes in G\"odel universes}

\vspace{48pt}

{\bf Cristian Stelea,~}
{\bf Kristin Schleich,~}
{\bf and Donald Witt}

\vspace*{0.2cm}

{\it Department of Physics and Astronomy, University of British Columbia}\\
{\it 6224 Agricultural Road, Vancouver, BC V6T 1Z1, Canada}\\[.5em]

\end{center}

\vspace{30pt}

\begin{abstract}
 We investigate five-dimensional rotating and charged black holes with squashed horizons in G\"odel universes. The general solution was recently derived by applying a squashing transformation on the general non-extremal charged and rotating black hole in the G\"odel universe found by Wu. We give a discussion of the squashed geometry and also consider its lift to ten dimensions and discuss the $T$-dual geometry. Finally, using the counterterms method we compute its conserved charges and explore its thermodynamics. 
\end{abstract}

\vspace{32pt}

\setcounter{footnote}{0}

\newpage

\section{Introduction}

In recent years,  one has witnessed remarkable developments in higher-dimensional black hole physics. Even though several higher-dimensional black holes have been known for a long time, such as the static Schwarzschild-Tangherlini black holes, their charged  Reissner-Nordstr\"om versions and the higher dimensional generalization of the rotating Kerr solution \cite{Tangherlini:1963bw,Myers:1986un}, one of the most intriguing recent results concerns the existence of higher dimensional asymptotically flat black holes with non-spherical horizon topology. In particular, in five dimensions there exists an asymptotically flat black ring solution whose horizon's topology is $S^2\times S^1$ instead of the usual $S^3$ topology of the Schwarzschild solution \cite{Emparan:2001wn}. One can imagine constructing such a black ring by taking a black string in five-dimensions, bending it and connecting its ends to form a circle. This configuration would normally be expected to collapse  to form a black hole with spherical horizon topology;
 indeed, this is the case in four dimensional asymptotically flat spacetimes  as a consequence of topological censorship \cite{Friedman:1993ty,Galloway:1999bp}. However, in five or more dimensions, the spherical topology of infinity does not constrain that of the black hole horizon \cite{Galloway:1999br}; geometric considerations, however, restrict the topology to those, such as $S^3$ and $S^2\times S^1$, that admit non-negative scalar curvature \cite{Cai:2001su}.
 It was found that if one adds rotation, the black ring can be stabilized  \cite{Emparan:2001wn}; this provided the first known explicit solution with non-spherical horizon in an asymptotically flat background. One can also envisage that adding a gauge field can also stabilize the gravitational collapse of the black ring. Such a solution has been explicitly constructed in the five-dimensional Einstein-Maxwell theory \cite{Ida:2003wv,Kunduri:2004da,Yazadjiev:2005hr,Yazadjiev:2005gs}. However, it was found that even by adding an electric charge, the static black ring cannot be stabilized. Supersymmetric black rings have also been found \cite{Elvang:2004rt, Gauntlett:2004wh} based on the procedure given in \cite{Gauntlett:2002nw}.

In five dimensions there also exist the so-called Kaluza-Klein (KK) black holes, whose horizon geometry is a squashed $3$-sphere \cite{Dobiasch:1981vh,Gibbons:1985ac}. However, their geometry is not asymptotically flat; instead it is asymptotic to a nontrivial $S^1$ bundle with constant fiber over the 2-sphere in a four dimensional asymptotically flat spacetime. This is also the asymptotic geometry of the Kaluza-Klein monopole \cite{Sorkin:1983ns,Gross:1983hb}. The Kaluza-Klein monopole, one of the most remarkable geometries, is a solution of the five-dimensional vacuum Einstein theory. It is perfectly regular and yet, upon dimensional reduction to four dimensions, one obtains the geometry of a magnetic monopole, which is nakedly singular from the four-dimensional point of view. The KK-monopole geometry is characterized by the radius of the  Kaluza-Klein circle $S^1$. For distances much smaller than this radius, the geometry becomes isometric to the five-dimensional Minkowski geometry, while asymptotically the geometry looks  essentially four-dimensional, as the Kaluza-Klein circle is small. This property of the KK monopole background, that it  interpolates between a five-dimensional geometry and an essentially four-dimensional one, has been extensively used in context of supergravity theories. For instance, given a supersymmetric solution, one can add straightforwardly a KK-monopole charge and therefore obtain a new four-dimensional solution upon dimensional reduction. This has led to remarkable connections between five-dimensional supersymmetric black holes and certain four-dimensional black holes \cite{Gaiotto:2005gf,Elvang:2005sa}.

 For vacuum metrics, there exists a systematic procedure to add KK-monopole charge to a general asymptotically flat geometry \cite{Ford:2007th}, based on a hidden $SL(3,R)$ symmetry of the gravitational sector \cite{Maison:2000fj}. More recently, in the context of the minimal five-dimensional supergravity, there has been developed a solution generating technique based on the $G_2$ U-duality arising in the dimensional reduction of the theory down to three dimensions \cite{Bouchareb:2007ax,Clement:2007qy}.  One should note here that adding a Kaluza-Klein monopole charge to a given solution  is not a trivial task once we depart from the class of supersymmetric/vacuum solutions;  in many cases, to find exact solutions one has to solve the Einstein-Maxwell equations by brute force. For instance, a solution describing a static KK black hole with electric charge has been found in \cite{Ishihara:2005dp}, while the corresponding Einstein-Yang-Mills solution has been described in \cite{Brihaye:2006ws}. The Ishihara-Matsuno solution has been further generalized to a solution of minimal supergravity in five-dimensions in \cite{Nakagawa:2008rm}. Remarkably, with hindsight, such KK solutions can be generated by applying a `squashing' transformation on suitable asymptotically flat geometries. This `squashing' transformation was initially used in \cite{Wang:2006nw} to obtain the vacuum rotating black hole with squashed horizons starting from the five-dimensional Kerr solution with equal rotation parameters. Since it works on asymptotically flat solutions of the minimal supergravity in five dimensions, the effects of the squashing transformation should presumably be recovered and further extended in the context of the $G_2$ solution generating technique. However, the parameterization of the scalar cosets in this method is currently too complicated to be used as an effective solution generating technique. It is therefore of interest to check the effects of the `squashing' transformation on a case-by-case basis on various other solutions of the minimal supergravity in five-dimensions in order to see if one can generate new exact solutions. 
 
 One such important class of solutions corresponds to the so-called G\"odel-type solutions. In four dimensions an exact solution describing a rotating universe was found by G\"odel \cite{Godel:1949ga}. The G\"odel solution is a homogeneous rotating solution of Einstein's equations with pressureless matter and negative cosmological constant and, historically, it was one of the first exact solutions of Einstein's equations to exhibit close timelike curves (CTCs) through every point. Supersymmetric generalizations of the G\"odel universe in five dimensions have been found in  \cite{Gauntlett:2002nw}. Just as in the original four-dimensional G\"odel solution, they exhibit CTCs through every point. Further exact solutions describing various black holes (supersymmetric or not) embedded in a G\"odel universe have been found \cite{Herdeiro:2002ft,Gimon:2003ms,Herdeiro:2003un, Brecher:2003wq, Behrndt:2004pn, Wu:2007gg} and it turns out that, by applying the `squashing' transformation it is possible to generate new `squashed' black hole solutions \cite{Tomizawa:2008hw,Matsuno:2008fn,Tomizawa:2008rh}.

The purpose of this paper is specifically to investigate the properties of the new solution obtained by applying the squashing transformation to the general non-extremal charged black hole embedded in the G\"odel universe found by Wu in \cite{Wu:2007gg}.

The structure of this paper is as follows: in the next section we describe the squashed G\"odel black hole and show how previous solutions are recovered in the various limits. We then go on to check in detail the effect of the squashing factor on the background G\"odel geometry. In section $4$ we describe the lift to eleven dimensions and the $T$-dual geometry. Using the counterterms method, we compute the conserved quantities and discuss some  thermodynamic properties of the general solution. We find that, as expected, the conserved quantities satisfy a generalized Smarr relation and that the  first law of black hole thermodynamics has to modified to account for the effects of the gravitational tension and the dipole charge. Unexpectedly, there are ranges of the parameters for which some of the intriguing thermodynamical properties of the black holes in G\"odel universes are preserved: in particular we found that the black hole entropy is bounded from above and it can decrease while one increases the mass of the black hole.  We end with a summary of our work and consider avenues for future research.

\section{The solution}

The bosonic sector of the minimal supergravity in five dimensions is described by the Einstein-Maxwell Lagrangian with Chern-Simons terms:
\beqs
{\cal L}&=&\frac{1}{16\pi G}\bigg[\sqrt{-g}(R-F^2)-\frac{2}{3\sqrt{3}}\epsilon^{\alpha\beta\gamma\mu\nu}A_{\alpha}F_{\beta\gamma}F_{\mu\nu}\bigg].
\eeqs
The equations of motion derived from this Lagrangian are given by:
\beqs
R_{\mu\nu}&=&2\left(F_{\mu\alpha}F^{\alpha}_{\nu}-\frac{1}{6}g_{\mu\nu}F_{\alpha\beta}F^{\alpha\beta}\right),\nonumber\\
D_{\mu}F^{\mu\nu}&=&\frac{1}{2\sqrt{3}\sqrt{-g}}\epsilon^{\alpha\beta\gamma\delta\nu}F_{\alpha\beta}F_{\gamma\delta},
\eeqs
where $\epsilon^{\alpha\beta\gamma\delta\nu}$ is the Levi-Civita symbol.

To find a solution to the supergravity equations of motion and motivated by the `squashing transformation' introduced in \cite{Wang:2006nw} we used the following metric ansatz:
\beqs
ds^2&=&-k(r)dt^2-2g(r)\sigma_3dt+h(r)\sigma_3^2+\frac{\chi^2(r)}{V(r)}dr^2+\frac{r^2}{4}\bigg[\chi(r)(\sigma_1^2+\sigma_2^2)\bigg],\label{metricansatz}
\eeqs
while the ansatz for the electromagnetic $1$-form potential is
\beqs
A&=&B(r)dt+C(r)\sigma_3.\label{Avalues}
\eeqs
Here we used the left-invariant one forms on $S^3$, with Euler angles $(\theta,\phi,\psi)$:
\beqs
\sigma_1&=&\sin\psi d\theta-\cos\psi\sin\theta d\phi,\nonumber\\
\sigma_2&=&\cos\psi d\theta+\sin\psi\sin\theta d\phi,\nonumber\\
\sigma_3&=&d\psi+\cos\theta d\phi.
\eeqs
Substituting this ansatz into the equation of motion leads to the following general solution (see also \cite{Tomizawa:2008hw}):
\beqs
k(r)&=&1-\frac{2m}{r^2}+\frac{q^2}{r^4},\nonumber\\
g(r)&=&jr^2+3jq+\frac{(2m-q)a}{2r^2}-\frac{q^2a}{2r^4},\nonumber\\
h(r)&=&-j^2r^2(r^2+2m+6q)+3jqa+\frac{(m-q)a^2}{2r^2}-\frac{q^2a^2}{4r^4}+\frac{r^2}{4},\\
\chi(r)&=&\frac{c^2+2\left(m-4j(m+q)[a+2j(m+2q)]\right)c+q^2+2a^2(m-q)-8q^2j[2a+j(m+3q)]}{(r^2+c)^2},\nonumber\\
V(r)&=&1-\frac{2m}{r^2}+\frac{8j(m+q)[a+2j(m+2q)]}{r^2}+\frac{2(m-q)a^2+q^2[1-16ja-8j^2(m+3q)]}{r^4},\nonumber
\eeqs
while
\beqs
B(r)&=&\frac{\sqrt{3}q}{2r^2}-\Psi,~~~~~~~~ C(r)=\frac{\sqrt{3}}{2}\left(jr^2+2jq-\frac{aq}{2r^2}\right)\label{BCvalues},
\eeqs
where $\Psi$ is a constant to be fixed later by demanding a regular electromagnetic potential on the horizon.

Here $c$ is a real constant and it controls the squashing of the three sphere. In order to keep the signature of the metric Lorentzian, one has to enforce the condition $\chi(r)>0$. Let us note that $c$ can take any value, positive or negative as long as $\chi(r)$ remains positive. The solution corresponding to negative values of $c$ has been previously investigated in \cite{Tomizawa:2008hw}. The squashing effect disappears and one recovers the general non-extremal rotating black hole in G\"odel universe found in \cite{Wu:2007gg} in the limit $c\rightarrow\infty$. Its extremal version, which is supersymmetric has been studied in \cite{Herdeiro:2002ft}. If one sets $j=0$, one recovers the squashed charged black hole solution of the five-dimensional Einstein-Maxwell system with Chern-Simon terms \cite{Nakagawa:2008rm}.

Before we embark in a discussion of the squashed solutions, let us first recall some basic properties of the initial `un-squashed' solutions. For this purpose, it turns out that it is more convenient to recast the metric in an $ADM$-like decomposition:
\beqs
ds^2&=&-\frac{r^2V(r)}{4h(r)}dt^2+h(r)\left(\sigma_3-\frac{g(r)}{h(r)}dt\right)^2+\frac{\chi^2}{V(r)}dr^2+\frac{\chi r^2}{4}(\sigma_1^2+\sigma_2^2),\nonumber\\
&=&-\frac{\Delta}{r^4\beta}dt^2+\frac{r^2\beta}{4}\left(\sigma_3-\frac{4\tilde{g}}{r^4\beta}dt\right)^2+\frac{r^4\chi^2}{\Delta}dr^2+\frac{\chi r^2}{4}(\sigma_1^2+\sigma_2^2),
\eeqs
where now:
\beqs
\Delta&=&r^4-2[m-4j(m+q)[a+2j(m+2q)]]r^2+2(m-q)a^2+q^2[1-16ja-8j^2(m+3q)],\nonumber\\
\tilde{g}&=&jr^4+3jqr^2+\frac{(2m-q)a}{2}-\frac{q^2a}{2r^2},\nonumber\\
\beta&=&\frac{r^2\Delta-4\tilde{g}^2}{r^2(r^4-2mr^2+q^2)}.
\label{beta}
\eeqs
In general the solution is singular at origin $r=0$. The black hole horizons are located at the roots of $\Delta=0$:
\beqs
r_{H}^2&=&m-4j(m+q)(a+2j(m+2q))\pm\sqrt{\delta},\nonumber\\
\delta&=&(m-q-8j^2(m+q)^2)[m+q-2a^2-8ja(m+2q)-8j^2(m+2q)^2].
\eeqs
Here the plus/minus signs correspond to an outer/inner black hole horizon and  they  coincide in the extremal case, when the parameters are chosen such that $\delta=0$. The roots of $k(r)$ correspond to ergospheres. Explicitly, they are given by $r_e^2=m\pm\sqrt{m^2-q^2}$. If the charge parameter is set to zero, $q=0$, then there is only one ergosphere located at $r_e=\sqrt{2m}$.

Generically, as was first noticed in \cite{Cvetic:2005zi}, the gauge potential is not regular at the horizons and one finds that $A_{\mu}A^{\mu}$ diverges for $r=r_H$, unless one takes:\footnote{From here on we shall designate by $B(r)=\frac{\sqrt{3}q}{2r^2}$, without the constant term.}
\beqs
\Psi&=&B(r_{H})+\frac{4C(r_{H})\tilde{g}(r_{H})}{\beta(r_{H})r_H^4}\equiv B(r_{H})+\frac{C(r_{H})g(r_{H})}{h(r_{H})} .
\label{elpot}
\eeqs

If the condition $\beta>0$ is violated then the Killing vector $\partial_{\psi}$  becomes timelike;as $\psi$ is periodic, the metric then admits closed timelike curves in the region where $\beta<0$. This region has been referred to as the `time machine' in \cite{Cvetic:2005zi}. Its boundary is the so-called `velocity of light surface' (VLS) and corresponds to locations $r=r_{CTC}$, where $r_{CTC}$ is a root of $\beta=0$. 

One can also see explicitly that VLS cannot be situated in between the black hole horizons $r_{H}^{\pm}$. Indeed, from the general form of the function $\beta$ as given in (\ref{beta}) one notices that the only way to achieve $\beta(r_{CTC})=0$ is to have $\Delta(r_{CTC})\geq 0$, while if $r_{CTC}$ is in between the black hole horizons, $r_{H}^{-}<r_{CTC}<r_H^+$, one has $\Delta<0$. In the uncharged case, $q=0$, this fact has been explicitly checked in \cite{Kerner:2007jk}.

The angular velocity of locally non-rotating observers is given by the shift vector and, when evaluated on the horizon, it gives the angular velocity of the horizon:
\beqs
\Omega_H&=&\frac{4\tilde{g}(r_H)}{r_H^4\beta(r_H)}\equiv \frac{g(r_H)}{h(r_H)}.
\eeqs
For a specific choice of parameters, one should be able to make the angular velocity of the horizon vanish. This, however, can happen only in the degenerate case in which $\tilde{g}(r_H)=0$, \textit{i.e.} $\Delta_{r_H}=\beta_{r_H}=0$, and one finds that the VLS coincides with one of the black hole horizons.

Therefore, one has two important cases to consider: in the first case the VLS is inside the inner black hole horizon, while in the second, the VLS is outside the outer black hole horizon. In the second case it turns out that inside the time machine, the coordinate $t$ is spacelike and the spacetime  ends where the coefficient of $dt^2$ vanishes. This corresponds to the so-called `pseudo-horizon' \cite{Cvetic:2005zi}. To make the geometry non-singular at this point, one has to periodically identify the coordinate $t$ with an appropriate real period.

\section{Properties of the squashed geometry}

One is now ready to discuss the effects of the squashing function $\chi(r)$ on the initial undeformed geometry. There are two cases to consider, depending on the sign of the squashing parameter $c$.

If $c<0$, one can write $c=-r_0^2$ for some positive $r_0$ and, in this case, notice that $\chi(r)=\frac{\Delta(r_0)}{(r^2-r_0^2)^2}$. In order to keep the Minkowskian signature of the spacetime after the introduction of the squashing factor, one has to make sure that $\Delta(r_0)>0$, \textit{i.e.} $r_0$ cannot be located in between the black hole horizons (where $\Delta(r_0)<0$) and therefore $r_0>r_H^+$ or $r_0<r_H^-$. On the other hand, if $c>0$ then one can write $c=r_0^2$ and $\chi(r)=\frac{\Delta(ir_0)}{(r^2+r_0^2)^2}$. The constraint $\chi(r)>0$ translates in this case to $\Delta(ir_0)>0$ and, given that $\Delta$ is a quadratic function of $r_0^2$, one finds again that one cannot pick $r_0$ to lay in between the black hole horizons.

\subsection{The squashed G\"odel background}

Let us first investigate the squashing effects on the background G\"odel geometry, $m=a=q=0$. If $c<0$ one finds $\chi(r)=\frac{r_0^4}{(r^2-r_0^2)^2}$, which diverges at $r=r_0$. However, this is not a real singularity in the metric, as no curvature invariants diverge there. Moreover, it can be shown by an appropriate coordinate transformation that the region around $r_0$ can be considered as the region near spatial infinity \cite{Ishihara:2005dp}. Since there are no event horizons in the background geometry, one can choose any real value for $r_0$. There exists a VLS located at  $\beta(r_{CTC})=0$, \textit{i.e.} at $r_{CTC}=\frac{1}{2j}$ and for $r_0>r_{CTC}$ there will still be CTC's through every point in the squashed geometry. If, however, one picks $r_0<r_{CTC}$ then CTC are excluded from the region $r<r_0$.

One should note here that the limit $r\rightarrow r_0$ can be attained from the left or from the right of $r_0$. In the first case, one considers only values $r<r_0$ and performs the coordinate change:
\beqs
r^2&=&\frac{r_0^2}{1+\frac{\rho_0}{\rho}}\equiv \frac{r_0^2}{f(\rho)}.
\label{lefttransf}
\eeqs
Clearly, this coordinate change is only valid in the region $r<r_0$. In the limit $r\rightarrow r_0$ one has $\rho\rightarrow\infty$. So far, the constant $\rho_0$ has been left arbitrary. One can take $r_0^2=4\rho_0^2$, while $j$ can be rescaled to absorb the extra factor $r_0^2$. Then the squashed metric can then be written in the form:
\beqs
ds_L^2&=&-\left(dt+\frac{j}{f(\rho)}\sigma_3\right)^2+f(\rho)d\rho^2+\rho^2f(\rho)(\sigma_1^2+\sigma_2^2)+\frac{\rho_0^2}{f(\rho)}\sigma_3^2,\nonumber\\
A_L&=&\frac{\sqrt{3}}{2}\frac{j}{f(\rho)}\sigma_3.\label{leftgodel}
\eeqs
The geometry is everywhere non-singular and there is no ergoregion (since $g_{tt}=-1$ everywhere).\footnote{Note that in \cite{Tomizawa:2008hw}, the existence of ergoregions was identified in coordinates  which are not rotating  at infinity.} If $j=0$ the metric reduces to that of a Kaluza-Klein monopole. Notice that $g_{\psi\psi}=\frac{\rho_0^2f(\rho)-j^2}{f^2(\rho)}$ and, therefore, there exists a VLS located at $\rho_{CTC}=\frac{\rho_0^3}{j^2-\rho_0^2}$ iff $j^2>\rho_0^2$. The limit $\rho_0\ra 0$ can be taken once we further shift the $\rho$ coordinate $\rho\ra \rho-\frac{\rho_0}{2}$ and make the rescalings $\psi\ra \frac{\psi}{\rho_0}$ and $j\ra j\rho_0$ such that the metric becomes (with $\frac{\rho_0}{2}\equiv n$):
\beqs
ds_L^2&=&-\left(dt+j\frac{\rho-n}{\rho+n}(d\psi+2n\cos\theta d\phi)\right)^2+\frac{\rho+n}{\rho-n}d\rho^2+(\rho^2-n^2)(d\theta^2+\sin^2\theta d\phi^2)\nonumber\\&&+\frac{\rho-n}{\rho+n}(d\psi+2n\cos\theta d\phi)^2,\nonumber\\
A_L&=&\frac{\sqrt{3}}{2}j\frac{\rho-n}{\rho+n}(d\psi+2n\cos\theta d\phi).
\eeqs
For $n=0$, one recovers essentially  flat spacetime in boosted coordinates, while for $j=0$ one recovers the metric of the KK-monopole in Taub-NUT like coordinates. The limit $\rho_0\ra\infty$ can also be achieved after one rescales $\rho\ra\frac{\rho}{\rho_0}$ and $j\ra j\rho_0^2$, in which case one obtains:
\beqs
ds_L^2&=&-\left(dt+j\frac{\rho\rho_0^2}{\rho+\rho_0^2}\sigma_3\right)^2+\frac{\rho+\rho_0^2}{\rho\rho_0^2}d\rho^2+\frac{\rho(\rho+\rho_0^2)}{\rho_0^2}(\sigma_1^2+\sigma_2^2)+\frac{\rho\rho_0^2}{\rho+\rho_0^2}\sigma_3^2,\nonumber\\
A_L&=&\frac{\sqrt{3}}{2}j\frac{\rho\rho_0^2}{\rho+\rho_0^2}\sigma_3.
\label{2in1}
\eeqs
It is now clear that in the limit $\rho_0\ra\infty$, the squashing disappears and one recovers the initial G\"odel solution, with $\rho=r^2/4$.

On the other hand, if one approaches the limit $r\rightarrow r_0$ from the right, then the appropriate coordinate transformation is given by:
\beqs
r^2&=&\frac{r_0^2}{1-\frac{\rho_0}{\rho}}\equiv \frac{r_0^2}{\tilde{f}(\rho)}.
\label{righttransf}
\eeqs
The limit $r\ra r_0$ is equivalent to $\rho\ra\infty$ and one obtains the metric:
\beqs
ds_R^2&=&-\left(dt+\frac{j}{\tilde{f}(\rho)}\sigma_3\right)^2+\tilde{f}(\rho)d\rho^2+\rho^2\tilde{f}(\rho)(\sigma_1^2+\sigma_2^2)+\frac{\rho_0^2}{\tilde{f}(\rho)}\sigma_3^2,\nonumber\\
A_R&=&\frac{\sqrt{3}}{2}\frac{j}{\tilde{f}(\rho)}\sigma_3,
\eeqs
where again we have rescaled $j$ to absorb a constant factor $r_0^2$ and set $r_0^2=4\rho_0^2$. If $j=0$, this metric reduces to that of a Kaluza-Klein monopole with negative KK monopole charge and it is therefore singular at $\rho=\rho_0$. One should note, however, that this region corresponds to $r\ra\infty$ in the original G\"odel background. The location of the VLS is given by the equation $\rho_0^2\tilde{f}(\rho_{CTC})-j^2=0$, that is $\rho_{CTC}=\frac{\rho_0^3}{\rho_0^2-j^2}$ if $\rho_0^2>j^2$. It is then clear that if the `left limit' metric has CTCs then the `right limit' metric does not and vice-versa.

Finally, if $c>0$ then the squashed metric becomes:
\beqs
ds^2&=&-\left(dt+jr^2\sigma_3\right)^2+\frac{c^8dr^2}{(r^2+c^2)^4}+\frac{r^2}{4}\bigg[\frac{c^4}{(r^2+c^2)^2}(\sigma_1^2+\sigma_2^2)+\sigma_3^2\bigg],\nonumber\\
A&=&\frac{\sqrt{3}}{2}jr^2\sigma_3.
\label{cpos}
\eeqs
This metric has no horizons, there is no curvature singularity at origin, however there exist curvature singularities at infinity, in the time machine region $r>r_{CTC}=\frac{1}{2j}$.

\subsection{The general squashed black hole}

The general case can be dealt with in the same manner. Given the presence of the black hole horizons, one cannot chose the parameter $r_0$ at will and, instead, one has to locate it inside or outside the black hole horizons.

Again we have two cases to consider, depending on the sign of the constant $c$ in the squashing factor. Since for a positive value of $c$ the background (\ref{cpos}) has un-interesting asymptotics, we shall focus in what follows on negative values of $c$. Consider then $c=-r_0^2$, for some positive $r_0$. The squashing factor $\chi(r)$ is then singular at $r=r_0$, however, as for the pure G\"odel case, this is a coordinate singularity. In fact, the region near $r_0$ corresponds to spatial infinity after one performs an appropriate change of coordinates. Notice that again, the limit $r\ra r_0$ can be reached either from the left or from the right of $r_0$, \textit{i.e.} using values $r<r_0$, respectively $r>r_0$. For each limit one obtains two distinct geometries, that we shall designate by `left' and `right' solution. The `left' solution is obtained by performing the coordinate change (\ref{lefttransf}):
\beqs
ds_L^2&=&-k(\rho)dt^2-2g(\rho)(d\psi+\cos\theta d\phi)dt+h(\rho)(d\psi+\cos\theta d\phi)^2+\frac{4\rho_0^2}{r_0^2}\frac{f(\rho)}{V(\rho)}d\rho^2\nonumber\\
&&+\rho^2f(\rho)(d\theta^2+\sin^2\theta d\phi^2),\nonumber\\
A_L&=&(B(\rho)-\Psi)dt+C(\rho)(d\psi+\cos\theta d\phi).
\label{leftsq}
\eeqs
where:
\beqs
f(\rho)&=&1-\frac{\rho_0}{\rho},\nonumber\\
k(\rho)&=&1-\frac{2m}{r_0^2}f(\rho)+\frac{q^2}{r_0^4}f^2(\rho),\nonumber\\
g(\rho)&=&j\frac{r_0^2}{f(\rho)}+3jq+\frac{(2m-q)a}{2r_0^2}f(\rho)-\frac{q^2a}{2r_0^4}f^2(\rho),\nonumber\\
h(\rho)&=&-j^2\frac{r_0^2}{f(\rho)}\left(\frac{r_0^2}{f(\rho)}+2m+6q\right)+3jqa+\frac{(m-q)a^2}{2r_0^2}f(\rho)-\frac{q^2a^2}{4r_0^4}f^2(\rho)+\frac{r_0^2}{4f(\rho)},\\
V(\rho)&=&1-\frac{2m-8j(m+q)[a+2j(m+2q)]}{r_0^2}f(\rho)+\frac{2(m-q)a^2+q^2[1-16ja-8j^2(m+3q)]}{r_0^4}f^2(\rho),\nonumber
\label{functions}
\eeqs
while
\beqs
B(\rho)&=&\frac{\sqrt{3}q}{2r_0^2}f(\rho),~~~~~~~~ C(\rho)=\frac{\sqrt{3}}{2}\left(j\frac{r_0^2}{f(\rho)}+2jq-\frac{aq}{2r_0^2}f(\rho)\right).
\eeqs
In the above form of the metric, we have already imposed the condition $V(r_0)=4\rho_0^2/r_0^2$ in order to have the canonical normalization of the metric on the $2$-sphere at infinity. However, the location of $r_0$ is still arbitrary at this stage.

Recall now that the general rotating charged black hole has generically two horizons situated at $r_H^{\pm}$ and  a VLS located at $r_{CTC}$. Since the VLS cannot be located in between the horizons one is left in general with two cases to examine; $r_{CTC}>r_H^+$ and $r_{CTC}<r_H^-$. If $r_{CTC}>r_H^+$ then if one picks $r_0$ such that $r_H^+<r_0<r_{CTC}$ with $h(r_0)>0$, then in consequence, $h(\rho)$ is always positive in the region corresponding to $r<r_0$ and there are no CTCs in the `left' geometry. On the other hand, if one considers the limit $r\ra r_0$ from the right, the appropriate coordinate change is given in (\ref{righttransf}) and the `right' geometry has the same form as (\ref{leftsq}) but with $f(\rho)$ replaced by $\tilde{f}(\rho)$, \textit{i.e.} one replaces $\rho_0\ra -\rho_0$. If the `left' geometry is regular, then the `right' geometry has no horizons and contains a naked singularity at $\rho=\rho_0$. The VLS in this case separates the region with CTCs from the asymptotic region with no CTCs.

If $r_0<r_H^-<r_H^+<r_{CTC}$ the `left' geometry has a naked singularity. However, the `right' geometry is now regular with an event horizon located at $\rho_H^-=\rho_0\frac{r_H^{-~2}}{r_H^{-~2}-r_0^2}$ with the VLS located inside the black hole horizons. Therefore there are no CTCs outside the horizon for this case. Finally, if the VLS in the un-squashed geometry is located inside the black hole horizons, then either $r_0<r_{CTC}<r_H^-<r_H^+$
 or $r_{CTC}<r_0<r_H^-<r_H^+$.  For each  of these cases, there is only one limit. Moreover, from the previous analysis, it is clear that  both the `left' and `right' geometries appear to be singular.

Finally, let us mention that it is also possible to rewrite the squashed solution in a form  from which one can recover the initial unsquashed geometry,  in a form similar to the metric (\ref{2in1}). To this end, one performs the rescaling $\rho\ra\frac{\rho \rho_0}{r_0^2}$ such that:
\beqs
ds^2_L&=&-k(\rho)dt^2-2g(\rho)(d\psi+\cos\theta d\phi)dt+h(\rho)(d\psi+\cos\theta d\phi)^2+\frac{V(r_0)^2}{4r_0^2}\frac{f(\rho)}{V(\rho)}d\rho^2\nonumber\\
&&+\frac{V(r_0)}{4r_0^2}\rho^2f(\rho)(d\theta^2+\sin^2\theta d\phi^2),\nonumber\\
A_L&=&(B(\rho)-\Psi)dt+C(\rho)(d\psi+\cos\theta d\phi),
\eeqs
where now $f(\rho)=1+\frac{r_0^2}{\rho}$, while the other functions remain the same as in (\ref{functions}). In the limit $r_0\ra\infty$, one recovers the initial unsquashed black hole metric. However,  the presence of the $V(r_0)$ factor multiplying the line element of the two-sphere leads to unnecessary complications in discussing the solution's properties. Therefore the next section we shall use the solution as given in (\ref{leftsq}).

\section{$T$-duality of the squashed G\"odel universe}

It is well known  that  five dimensional supergravity solutions can be extended  to 11 dimensional supergravity solutions of  the Lagrangian whose bosonic sector is
\beq
L =  \frac 1{2\kappa^2}  \sqrt{-g} (R- \frac 1{48} G_{\mu\nu\rho\sigma}G^{\mu\nu\rho\sigma}) - \frac 1{12\kappa^2} C_{(3)}\wedge G_{(4)}\wedge G_{(4)} \label{11lagrangian}
\eeq
where $C_{(3)}$ is the 3-form potential for $G_{(4)}$, $ G_{(4)} = dC_{(3)}$ and $\kappa^2$ is the 11 dimensional gravitational coupling constant.
The equations of motion for (\ref{11lagrangian}) are satisfied by the metric
\beq ds^2_{11} = g_{\mu\nu}dx^\mu dx^\nu + ds^2_{T^6}\label{11metric}\eeq where $g$ is the metric (\ref{metricansatz}),
\beq ds^2_{T^n} = \sum_{i=1}^n dy^2_i \,\eeq
the flat metric on the n-dimensional torus
and $C_{(3)}$ is chosen in a gauge such that
\beq C_{(3)} = \frac 2{\sqrt{3}} A\wedge K_{(2)}\eeq
where $A$ is the vector potential (\ref{Avalues}) and
\beq 
K_{(2)} =dy_1\wedge dy_2 + dy_3\wedge dy_4 + dy_5\wedge dy_6 
\eeq 
is the K\"ahler form on $T^6$.
Dimensional reduction of this solution along $y_6$  yields the type IIA string background
\beqs ds^2_{IIA} &=& g_{\mu\nu}dx^\mu dx^\nu + ds^2_{T^4}+ dz^2\label{IIAmetric}\cr
H_{(3)} &=& dB_{(2)} \eeqs
where $z=y_5$ is the coordinate paired with $y_6$ in (\ref{11metric}) and $B_{(2)}$ is given by
\beq B_{(2)}=\frac 2{\sqrt{3}} A\wedge dz =  \frac 2{\sqrt{3}} ( B(r)-\Psi) dt\wedge dz + \frac 2{\sqrt{3}} C(r) \sigma_3\wedge dz \eeq
where again $A$ is (\ref{Avalues}) with the convention that $B(r)=\frac{{\sqrt 3}q}{2r^2}$ as adopted in footnote 1.  $G$ for the IIA background is simply the projection of  the 4-form $G_{(4)}$ to the  10-dimensional spacetime. This spacetime is clearly  the generalization of the G\"odel IIA spacetime \cite{Boyda:2002ba} and the  G\"odel IIA black hole spacetime \cite{Gimon:2003ms} to the squashed, rotating, charged case.
A T-duality transformation of (\ref{IIAmetric}) along the $z$ direction\footnote{We follow the convention of  \cite{Harmark:2003ud} namely that the components of the IIB metric $\tilde g$ are related to the IIA metric $g$ and 2-form field $B$ by
$ \tilde g_{\mu\nu} = g_{\mu\nu} - g_{\mu z}g_{\nu z} + B_{\mu z} B_{\nu z}$
$\tilde g_{zz}=1$, $ \tilde g_{\mu z}=B_{\mu z}$ with the nontrivial RR and NS fields related by
$\tilde G_{\mu\nu\rho\sigma z} = G_{\mu\nu\rho\sigma }$, $\tilde F_{\mu\nu} = - H_{\mu\nu z}$.}
yields
 a new IIB string background:
\beqs ds^2_{IIB} &=& -\kappa(r) dt^2  +\frac 4{\sqrt{3}} (B(r)-\Psi) dt dz + dz^2 - 2\gamma(r)dt \sigma_3 +\eta(r) \sigma^2_3 +\frac 4{\sqrt{3}} C(r) dz \sigma_3 \cr && \hskip .1in + \frac {\chi^2(r)}{V(r)} dr^2 + \frac {r^2}{4} \chi(r)(\sigma_1^2 + \sigma_2^2) + ds^2_{T^4}\label{IIBmetric}
\eeqs where
\beqs \kappa(r) &=& 1-\frac{2 m}{r^2}+\frac{4 q \Psi }{\sqrt{3} r^2}-\frac{4 \Psi ^2}{3}\cr
\gamma(r)&=&2 j q+\frac{4 j q \Psi }{\sqrt{3}}+r^2 \left(j+\frac{2 j \Psi }{\sqrt{3}}\right)+\frac{a m-\frac{a q}{2}-2 j q^2-\frac{a q \Psi }{\sqrt{3}}}{r^2}\cr
\eta(r)&=& a j q+4 j^2 q^2+\frac{\frac{a^2 m}{2}-\frac{a^2 q}{2}-2 a j q^2}{r^2}+\left(\frac{1}{4}-2 j^2 m-2 j^2 q\right) r^2 \ .
\eeqs
Due to the form of $B_{(2)}$, the  $g_{tt} $, $g_{t\sigma_3}$ and $g_{\sigma_3\sigma_3}$ components now differ from that of the IIA metric  by $q$ and $\Psi$ dependent terms. 

It is interesting to consider the special case corresponding to the squashed G\"odel metric, $j\neq0$,  $a=m=q=\Psi=0$. First consider the left limit case; one finds, after coordinate transformation (\ref{lefttransf}), that the metric  (\ref{IIBmetric}) takes the form
\beqs ds^2_{L} &=& - dt^2+ dz^2 +2jr_0^2(1-\frac {\rho_0}{\rho+\rho_0})  \sigma_3 (dt-dz) +\frac {r_0^2}{4}(1-\frac {\rho_0}{\rho+\rho_0}) \sigma^2_3\cr && \hskip .1in  +\frac {r_0^2}{4\rho_0^2}(1+\frac {\rho_0}{\rho})\left(d\rho^2 +\rho^2(\sigma_1^2 + \sigma_2^2) \right)+ ds^2_{T^4}\ .
\eeqs
A time dependent angular coordinate transformation
$\psi'=\psi +4j(t-z)$
coupled to a change of coordinates
$u = t+z$, $v = t-z$
results in
\beqs ds^2_{L} &=& - dudv -4j^2r^2_0(1-\frac {\rho_0}{\rho+\rho_0})dv^2 +\frac {r_0^2}{4}(1-\frac {\rho_0}{\rho+\rho_0}) \sigma_3^2 \cr && \hskip .1in +\frac {r_0^2}{4\rho_0^2}(1+\frac {\rho_0}{\rho})\left(d\rho^2 +\rho^2(\sigma_1^2 + \sigma_2^2)  \right)+ ds^2_{T^4}\ \label{left limit}.
\eeqs
where now $\sigma_3 = d\psi' + \cos\theta d\phi$. This spacetime is a product of a 4-torus and  spacetime that is locally a product of an $S^1$ fiber  with a 5-dimensional spacetime. The $S^1$ fiber has constant radius at $\rho\to\infty$; The 5-dimensional spacetime becomes asymptotically flat as $\rho\to\infty$ and is everywhere regular.\footnote{Of course, this bundle structure assumes the appropriate periodic identification of $\psi'$.}   Regularity at $\rho = 0$ can be easily verified by expansion of (\ref{left limit})  in $\rho/\rho_0$ to leading order; the metric becomes
\beq d\bar s^2_{L} = - dudv +\frac {r^2_0}{\rho_0}\left(-4j^2\rho dv^2 +\frac {1}{4}\left(\frac {d\rho^2}{\rho}
+\rho \left(\sigma^2_1+\sigma^2_2+\sigma^2_3\right)\right)\right)+ ds^2_{T^4}\ \label{small rho limit}.
\eeq
The coordinate transformation $\rho=\frac{ \rho_0}{r_0^2}r^2$ then brings the spacetime near the origin to plane wave form
\beq d\bar s^2_{L} = - dudv -4j^2r^2 dv^2 + {dr^2}
+\frac{r^2}{4} \left(\sigma^2_1+\sigma^2_2+\sigma^2_3\right)+ ds^2_{T^4}\ \label{plane wave limit}.
\eeq  Note that  this metric becomes exact everywhere as $\rho_0\to\infty$, the limit of zero squashing. The zero squashing limit clearly coincides with the plane wave dual to G\"odel spacetime. Therefore the left limit metric can be described as a plane wave that asymptotes to a flat spacetime as $\rho\to\infty$. 

The right limit case can be constructed similarly; after the coordinate transformation (\ref{righttransf}) followed by the same coordinate transformations  as in the left limit case, metric (\ref{IIBmetric}) becomes
\beqs ds^2_{R} &=& - dudv -4r^2_0j^2(\frac {\rho}{\rho-\rho_0})dv^2 +\frac {r_0^2}{4}(\frac {\rho}{\rho-\rho_0}) \sigma_3^2 \cr && \hskip .1in +\frac {r_0^2}{4\rho_0^2}(1-\frac {\rho_0}{\rho})\left(d\rho^2 +\rho^2(\sigma^2_1 + \sigma^2_2)  \right)+ ds^2_{T^4}\ .
\eeqs
This spacetime has the same product structure as the left limit case.  The 5-dimensional base space again locally approaches flat spacetime as $\rho \to \infty$, corresponding to $r\to r_0$ in the original coordinates. However, in contrast to the left limit case, the spacetime is  singular as $\rho\to \rho_0$, that is $r\to \infty$ in the original coordinates.

The left limit of the general case yields
\beqs ds^2_{L} &=& -\kappa(\rho) dt^2  +\frac 4{\sqrt{3}} (B(\rho)-\Psi) dt dz + dz^2 - 2\gamma(\rho)dt \sigma_3 +\eta(\rho) \sigma^2_3 +\frac 4{\sqrt{3}} C(\rho) dz \sigma_3 \cr && \hskip .1in +\frac {r_0^2}{4\rho_0^2}(1+\frac {\rho_0}{\rho})\left(\frac{d\rho^2}{V(\rho)} +\rho^2(\sigma_1^2 + \sigma_2^2) \right) + ds^2_{T^4}\label{IIBleftmetric}\eeqs
where
\beqs \kappa(\rho) &=& 1-\frac{4 \Psi ^2}{3} -\left(\frac{2 m-\frac{4 q \Psi }{\sqrt{3}}}{r_0^2}\right)\left(\frac{\rho+\rho_0}{\rho}\right)\cr
\gamma(\rho)&=&2 j q+\frac{4 j q \Psi }{\sqrt{3}}+ \left(j+\frac{2 j \Psi }{\sqrt{3}}\right)r_0^2\left(\frac{\rho}{\rho+\rho_0}\right)+\left(\frac{a m-\frac{a q}{2}-2 j q^2-\frac{a q \Psi }{\sqrt{3}}}{r_0^2}\right)\left(\frac{\rho+\rho_0}{\rho}\right)\cr
\eta(\rho)&=& a j q+4 j^2 q^2+\left(\frac{\frac{a^2 m}{2}-\frac{a^2 q}{2}-2 a j q^2}{r_0^2}\right)\left(\frac{\rho+\rho_0}{\rho}\right)+\left(\frac{1}{4}-2 j^2 m-2 j^2 q\right) r_0^2\left(\frac{\rho}{\rho+\rho_0}\right)\cr
V(\rho)&=& 1-\left( \frac{2m-8j(m+q)[a+2j(m+2q)]}{r_0^2}\right)\left(\frac{\rho+\rho_0}{\rho}\right)+\cr && \hskip 1.5in
\left(\frac{2(m-q)a^2 + q^2[1-16ja-8j^2(m+3q)]}{r_0^4}\right)\left(\frac{\rho+\rho_0}{\rho}\right)^2 \cr
B(\rho)&=&\frac{\sqrt{3}q}{2r_0^2}\left(\frac{\rho+\rho_0}{\rho}\right)-\Psi \cr
C(\rho)&=&\frac{\sqrt{3}}{2} \left(2jq+ jr^2_0\left(\frac{\rho}{\rho+\rho_0}\right)-\frac{aq}{2r^2_0}\left(\frac{\rho+\rho_0}{\rho}\right)\right) \ .
\eeqs
For general parameters, the metric coefficients $\gamma$, $\eta$ and $C$ are not constant multiples of each other; therefore the metric cannot be brought into a form similar to that of a plane wave. Asymptotically, as $\rho \to \infty$, (\ref{IIBleftmetric}) takes the form
\beqs ds^2_{L} &=& -\kappa(\infty) dt^2  +\frac 4{\sqrt{3}} B(\infty) dt dz + dz^2 - 2\gamma(\infty)dt \sigma_3 +\eta(\infty) \sigma^2_3 +\frac 4{\sqrt{3}} C(\infty) dz \sigma_3 \cr && \hskip .1in +\frac {r_0^2}{4\rho_0^2}\left(\frac{d\rho^2}{V(\infty)} +\rho^2(\sigma_1^2 + \sigma_2^2) \right) + ds^2_{T^4}\eeqs
which can be written in a form similar to that of the asymptotic limit  of the IIB metric (\ref{left limit}) dual to the squashed G\"odel spacetime. 

In particular, observe that in the $q=a=0$ case, the 6-dimensional spacetime at constant $t,\rho,z$ 
has the form \beq ds^2 =\frac {r_0^2\rho^2}{4\rho_0^2}\left(\sigma_1^2 + \sigma_2^2\right)+\left(\frac{1}{4}-2 j^2 m\right) r_0^2\left(\frac{\rho}{\rho+\rho_0}\right)\sigma^2_3 \ . \eeq
In contrast to
\cite{Gimon:2003ms}, volume of the the 3-sphere scales as $\rho^2$ as $\rho\to\infty$ as the $S^1$ fiber approaches a constant radius in this limit.  Therefore, the squashed spacetime does not produce
an anisotropic 3-sphere in this limit, as was the case in \cite{Gimon:2003ms}; instead it becomes essentially a nontrivial $S^1$ bundle with constant fiber over an $S^2$ in flat spacetime. Hence, like the left IIA metric,  the general left IIB metric is also asymptotically of Kaluza-Klein black hole form.

\section{Conserved quantities and thermodynamics}

In discussing the thermodynamic properties of the charged rotating squashed black holes in the G\"odel universe we shall focus on the regular `left' geometry with parameters chosen such that $\delta\geq 0$ and $h(r_0)>0$. Moreover we also impose the condition that $r_0$ is outside the ergosphere, \textit{i.e.} $k(r_0)>0$.

The computation of  conserved charges and the discussion of the thermodynamic properties of black objects in G\"odel universes constitute a notoriously difficult task \cite{Boyda:2002ba,Klemm:2004wq,Barnich:2005kq} as the naive application of  traditional approaches fails. Fortunately, once the squashing factor is included, the geometry is essentially modified such that at large distances it becomes asymptotic to the KK-monopole background described in the previous section. For such backgrounds it turns out that one can consistently define conserved charges and perform a discussion of their thermodynamic properties.

In particular, to avoid the known problems of  background substraction methods, we shall recourse to the counterterms method as described in \cite{Mann:2005cx}.
Recall that the squashed G\"odel geometries are solutions of the equations of motion derived from the action:
\beqs
I&=&\frac{1}{16\pi G}\int_M d^5x\bigg[\sqrt{-g}(R-F^2)-\frac{2}{3\sqrt{3}}\epsilon^{\alpha\beta\gamma\mu\nu}A_{\alpha}F_{\beta\gamma}F_{\mu\nu}\bigg]+\frac{1}{8\pi G}\int_{\partial
M}K\sqrt{-h}\,d^{4}x.
\eeqs
Here $M$ denotes the bulk of a five-dimensional manifold, $\partial M$ denotes its boundary, while $K$ is the trace of the extrinsic curvature $K_{ij}=\frac{1}{2}h_{i}^{k}\nabla_{k}n_{j}$ of the boundary $\partial M$ with unit normal $n^{i}$ and induced
metric $h_{ij}$. The boundary Gibbons-Hawking term is required in order to obtain the Einstein equations upon applying a variational principle with metric variations but not their normal derivative fixed at the boundary.

Generically the action contains divergencies that arise from integrating over the infinite volume of spacetime. One way to regularize it for spacetime geometries that are asymptotic to that of a KK-monopole is to add the following surface counterterm:
\begin{equation}
I_{ct}=-\frac{1}{8\pi G}\int d^{4}x\sqrt{-h}\sqrt{2\mathcal{R}},  \label{I2}
\end{equation}
where $\mathcal{R}$ is the Ricci scalar of the induced metric on the
boundary, $h_{ij}$. By taking the variation of this total action with
respect to the boundary metric $h_{ij}$, it is straightforward to compute
the boundary stress-tensor, including (\ref{I2}):
\begin{equation}
T_{ij}=\frac{1}{8\pi G}\left( K_{ij}-Kh_{ij}-\Psi( \mathcal{R}_{ij}-\mathcal{R}h_{ij})-h_{ij}\Box\Psi+\Psi_{;ij}\right),
\end{equation}
where we denote $\Psi=\sqrt{\frac{2}{\mathcal{R}}}$.

If the boundary geometry has an isometry generated by a Killing vector $\xi ^{i}$, then $T_{ij}\xi ^{j}$ is divergence free, from which it follows that the quantity:
\begin{equation}
\mathcal{Q}=\oint_{\Sigma }d^{3}S^{i}T_{ij}\xi ^{j},
\end{equation}
associated with a closed surface $\Sigma $, is conserved. Physically, this
means that a collection of observers on the boundary with induced metric
$h_{ij}$ measure the same value of $\mathcal{Q}$, provided the boundary has
an isometry generated by $\xi $. In particular, if $\xi ^{i}=\partial
/\partial t$ then $\mathcal{Q}$ is the conserved mass $M$, while if $\xi^{i}=\partial/\partial \phi$ for some angular coordinate $\phi$ one obtains the angular momentum $J_{\phi}$. One should also note that for squashed Kaluza-Klein black holes there exists another conserved quantity, analogous to the tension in the black string case \cite{Kastor:2006ti,Harmark:2004ch}, which can be easily computed in the counterterms approach by using the formula:
\begin{equation}
\mathcal{T'}=\int_{\Sigma'}d^{3}S_{\psi}T^{\psi}_{j}\xi ^{j}=\int dt\oint_{S^2}d^2x\sqrt{\sigma}T^{\psi}_{\psi},
\end{equation}
where now $\xi ^{i}=\partial/\partial \psi$ and the integration is performed over the two-sphere at infinity (described by $\theta$ and $\phi$) and also along the time direction. This tension is defined with respect to the asymptotic spatial translation along the circle described by $\psi$. Similarly to the black string case, one
notices that one can omit the integration over time and work with the `tension per unit time':
\beqs
\mathcal{T}=\oint_{S^2}d^2x\sqrt{\sigma}T^{\psi}_{\psi},
\eeqs

Consider now the `left' geometry describing a regular rotating charged black hole. The event horizons are again located at the roots of $V(\rho)$ and they correspond to:
\beqs
\rho_{\pm}&=&\rho_0\frac{r_H^{\pm~2}}{r_0^2-r_{H}^{\pm~2}}.
\eeqs
The horizon topology is a squashed sphere with area
\beqs
{\cal A}_H&=&16\pi^2\rho_+^2f(\rho_+)\sqrt{h(\rho_+)}
\eeqs
and the associated Bekenstein-Hawking entropy is $S={\cal A}_H/4G$.
The angular velocity $\Omega_H$ of the horizon is:
\beqs
\Omega_H&=&\Omega_{\psi}=\frac{g(\rho_+)}{h(\rho_+)}.
\eeqs
To compute the mass, one needs an asymptotic timelike Killing vector $\xi=\partial/\partial t$ canonically normalized at infinity. Therefore,  for  metric (\ref{leftsq}), one cannot use  $\xi=\partial/\partial t$ directly  but  must normalize it such that $\xi_{\mu}\xi^{\mu}=-1$ at infinity. Further, one notices that the asymptotic frame in (\ref{leftsq}) is rotating with angular velocity:
\beqs
\Omega_{\infty}&=&\lim_{\rho\ra\infty}\frac{g(\rho)}{h(\rho)}
=\frac{4jr_0^6+12jqr_0^4+2a(2m-q)r_0^2-2q^2a}{r_0^6[1-4j^2(r_0^2+2m+6q)]+12jqar_0^4+2(m-q)a^2r_0^2-q^2a^2}.
\eeqs
A similar situation was encountered in the AdS/CFT context in computing the conserved quantities of asymptotically locally AdS backgrounds \cite{Caldarelli:1999xj,Papadimitriou:2005ii,Gibbons:2004ai}. The best way to avoid the subtleties in identifying the proper conserved mass and angular momentum is to perform directly a coordinate transformation to absorb the asymptotic angular velocity and, also, to rescale the time coordinate to obtain a canonically normalized timelike Killing vector:
\beqs
t\ra t/N_0,~~~~~~~\psi\ra \psi+\frac{\Omega_{\infty}}{N_0}t,
\eeqs
where $N_0^2=\lim_{\rho\ra\infty}\frac{r_0^2V(\rho)}{4h(\rho)}=\frac{r_0^2V_0}{4h_0}$ or:\footnote{Recall that we chose the parameter $r_0$ outside the event horizons/ergoregion and inside the VLS in order to avoid the presence of CTC's in the final squashed geometry. In this case all the square-roots in the formulae bellow in this section are well defined.} 
\beqs
N_0=\sqrt{\frac{r_0^6-2[m-4j(m+q)[a+2j(m+2q)]]r_0^4+2(m-q)a^2r_0^2+q^2[1-16ja-8j^2(m+3q)]r_0^2}{r_0^6[1-4j^2(r_0^2+2m+6q)]+12jqar_0^4+2(m-q)a^2r_0^2-q^2a^2}}.\nonumber\\
\eeqs

A straightforward computation using the boundary stress-tensor leads then to the following conserved mass:
\beqs
M&=&\frac{\pi}{16Gr_0^4h_0N_0}\bigg[r_0^6(1-8mj^2)(r_0^2+2m(1-8mj^2)-16mj^2r_0^2)-8mjar_0^6(1-8mj^2-8r_0^2j^2)\nonumber\\&&-4mr_0^2a^2\big[m(1-8mj^2)-12mj^2r_0^2-8r_0^4j^2\big]+16jr_0^2m^2a^3-4a^4m^2+q\big[-24r_0^6j^2(4m(1-8mj^2)\nonumber\\
&&+r_0^2(1-16mj^2))+16r_0^6ja(1+16j^2m+4r_0^2j^2)+4r_0^2a^2(m+16m^2j^2-8j^2r_0^4+24j^2r_0^2m)\nonumber\\
&&-48mjr_0^2a^3+8ma^4\big]+q^2\big[r_0^4(1280mr_0^2j^4-192m^2j^2+48mj^2+256r_0^4j^4-3-16r_0^2j^2)\nonumber\\
&&+16r_0^4ja(3-4r_0^2j^2-24mj^2)+a^2(2m-4r_0^2-144j^2r_0^4-16m^2j^2-32j^2r_0^2m)+16ja^3(m+2r_0^2)\nonumber\\
&&-4a^4\big]+q^3\big[48r_0^4j^2(3-24mj^2+8r_0^2j^2)-24r_0^2ja(1-8mj^2-48r_0^2j^2)+(320r_0^2j^2-64mj^2+2)a^2\nonumber\\
&&-48a^3j\big]+q^4\big[16j^2(36r_0^2ja-108j^2r_0^4-7a^2)\big]\bigg]
\label{masa}
\eeqs
 angular momentum:
\beqs
J_{\psi}&=&-\pi\frac{4r_0^6(2ja^2m+8aj^2m^2-am+6r_0^2j^2am+2r_0^6j^3)}{8r_0^6G}\\
&&-\pi\frac{2r_0^4q(24r_0^6j^3-3r_0^4j+6r_0^4j^2a+24r_0^4j^3m+40r_0^2j^2am-4r_0^2ja^2+r_0^2a-6ja^2m)}{8r_0^6G}\nonumber\\
&&-\pi\frac{r_0^4q^2(144r_0^4j^3-64aj^2r_0^2-6a^2j+3a-24j^2ma)-aq^3(72r_0^4j^2-12r_0^2ja+a^2)}{8r_0^6G},\nonumber
\eeqs
and gravitational tension:
\beqs
{\cal T}&=&-\frac{1}{64Gr_0^6\sqrt{h_0^3}N_0}\bigg[2r_0^2\big[8m^2r_0^2a^3(a+2r_0^2j)-4r_0^4ma^2[m(1-8mj^2)-2j^2r_0^4]-8mjar_0^8(1-8mj^2\nonumber\\
&&-8r_0^2j^2)+2r_0^8(64m^3j^4+64r_0^2m^2j^4+6j^2r_0^4+m-16m^2j^2-r_0^2)\big]+q\big[24a^3r_0^4mj-16mr_0^2a^4\nonumber\\
&&+4r_0^4a^2(+16m^2j^2-2j^2r_0^4)+4r_0^8ja(16r_0^2j^2+64mj^2-5)+48j^2r_0^8(r_0^2-2m+8mj^2r_0^2\nonumber\\
&&+16m^2j^2)\big]+q^2\big[4r_0^8j^2(64r_0^2j^2+344j^2m+32aj-7)-a^2r_0^2(32j^2r_0^2m-8m+r_0^2+64m^2j^2)\nonumber\\
&&-40r_0^2j(r_0^2+2m)a^3+(8r_0^2-6m)a^4\big]+q^3\big[672r_0^8j^4+12r_0^4aj(1-8mj^2)-4r_0^2a^2(1+40mj^2\nonumber\\
&&+64r_0^2j^2)+48a^3r_0^2j+6a^4\big]+q^4\big[48a^3j+(24mj^2-3+32r_0^2j^2)a^2-288j^3ar_0^4\big]+72q^5a^2j^2\bigg].\nonumber\\
\eeqs

As first noticed in \cite{Cvetic:2005zi}, the gauge field is generically singular on the horizon and one has to perform a suitable gauge transformation (this explains the presence of $\Psi$ in the gauge potential $A_{\mu}$) to make it regular. Indeed, in the coordinates given in (\ref{leftsq}) one has:
\beqs
A_{\mu}A^{\mu}&=&\frac{C^2(\rho)}{h(\rho)}-\frac{4f(\rho)[g(\rho)C(\rho)+h(\rho)(B(\rho)-\Psi)]^2}{r_0^2V(\rho)h(\rho)}.
\eeqs
On horizon $V(\rho_+)=0$ and the above quantity diverges unless one picks:
\beqs
\Psi&=&B(\rho_+)+\frac{g(\rho_+)}{h(\rho_+)}C(\rho_+).
\label{phihor}
\eeqs

On the other hand, the electric potential $\Phi$, which is the true scalar potential that appears in the first law of black hole thermodynamics and also in the Smarr relation, is generically measured at infinity with respect to the horizon and is therefore defined by:
\beqs
\Phi&=&A_{\mu}\chi^{\mu}|_{\rho\ra\infty}-A_{\mu}\chi^{\mu}|_{\rho=\rho_+},\nonumber\\
&=&\frac{\Psi-B(\rho)-C(\rho)\Omega_H}{N_0}|_{\rho\ra\infty},
\eeqs
where $\chi^{\mu}=\partial/\partial t + \Omega \partial/\partial \psi$ is the null generator of the horizon (with $\Omega=(\Omega_H-\Omega_{\infty})/N_0$) and $A_{\mu}$ is the gauge potential. 

The Euclidean section is achieved by sending $t\ra -i\tau$, $a\ra i\tilde{a}$ and $j\ra i\tilde{j}$. The Hawking temperature can now be computed by employing the usual Euclidean section techniques once we extend the metric smoothly onto the horizon at $\rho=\rho_+$ and assign $\tau$ a period $\beta=\frac{1}{T}$, with the result:
\beqs
T&=&\frac{\sqrt{h_0}(\rho_+-\rho_-)}{4\pi\rho_+^2f(\rho_+)\sqrt{h(\rho_+)}}.
\label{temp}
\eeqs

The Euclidean action can be evaluated easily once one makes use of the field equations to prove that $R=F^2/3$, while the $F^2$ action term can be integrated by parts and, using the equations of motion for the Maxwell field, the net result can be expressed as a difference of two boundary terms:
\begin{eqnarray}
-\int_M d^5x\sqrt{g} \Big( -\frac{2}{3}F^2  -
\frac{2}{3\sqrt{3}\sqrt{g}} F\wedge F\wedge A
\Big)&=&\frac{4}{3}(\oint_{\infty} d^4x
\sqrt{g}~n_{\mu}A_{\nu}F^{\mu \nu}- \oint_{H} d^4x
\sqrt{g}~n_{\mu}A_{\nu}F^{\mu \nu}).\nonumber\\
\end{eqnarray}
A direct computation reveals that the event horizon contribution in
the above relation vanishes but only after using the regularized gauge potential $A$ with $\Psi$ given by (\ref{phihor}). After taking into account the boundary counterterm and analytically continuing back the parameters $\tilde{j}$ and $\tilde{a}$ the action can then be written as:
\beqs
I_E&=&-\frac{\beta\sqrt{3}\pi}{6GN_0}\Psi \bigg[4j(ma+jr_0^4)+q(16j^2r_0^2-(1-8j^2m))+\frac{q^2}{r_0^4}(24j^2r_0^4+a^2-8ar_0^2j)\bigg]\nonumber\\
&&+\frac{\pi\beta}{4Gr_0^4N_0}\bigg[r_0^6(1-8j^2m)+4j^2r_0^8-2ma^2r_0^2+2qr_0^2(a^2-8mj^2r_0^2+4j^2r_0^4)
\nonumber\\&&+q^2(16ajr_0^2+8j^2r_0^4-a^2)+4jq^3(-a+4jr_0^2)\bigg],
\eeqs
The electric charge $Q$ is computed using the Gauss formula by taking into account the Chern-Simon contribution:
\beqs
Q&=&-\frac{1}{4\pi G}\int_{\Sigma}d\Sigma_{\mu\nu}\bigg[F^{\mu\nu}+\frac{1}{\sqrt{3}\sqrt{-g}}\epsilon^{\mu\nu \alpha\beta\gamma}A_{\alpha}F_{\beta\gamma}\bigg]\nonumber\\
&=&-\frac{\sqrt{3}}{2}\frac{\pi}{G}\big[-4jma+q(1-8mj^2+4ja)-8j^2q^2\big].
\eeqs
One can also define a magnetic dipole charge by :
\beqs
D&=&\frac{1}{4\pi}\int_{S^2}F=\frac{\sqrt{3}}{2}jr_0^2,
\label{dipolecharge}
\eeqs
where one integrates over the $2$-sphere at infinity. The role of the dipole charges in black hole thermodynamics has been discussed in \cite{Copsey:2005se} (see also \cite{Emparan:2004wy,Astefanesei:2005ad}). In terms of these quantities the first law of black hole thermodynamics can be written as:
\beqs
dM&=&TdS+\Phi dQ+{\cal T}d{\cal L}+\Omega dJ_{\psi}+\Phi_D dD
\label{firstlaw}
\eeqs
and there also exists a generalized Smarr relation of the form:
\beqs
2M&=&3TS+\Phi Q+{\cal T}{\cal L}+\Phi_D D+3\Omega J_{\psi}.
\label{smarr}
\eeqs
Here ${\cal L}$ is the length of the $S^1$ circle at infinity described by $\psi$ and the above version of first law takes now explicitely into account variations of this length. The presence of the gravitational tension term in the first law of thermodynamics for the squashed black holes has been anticipated in \cite{Cai:2006td} and later confirmed in \cite{Kurita:2007hu,Kurita:2008mj} although not exactly in the simple form given above. 

Note that the five-dimensional Smarr relation derived in \cite{Gauntlett:1998fz} for asymptotically flat spacetimes has to be modified once one takes into account the effects of the dipole charge $D$ respectively of the gravitational tension ${\cal T}$. At this stage we shall not prove the general Smarr relation given above, however we shall verify it in particular case at the end of this section.
Notice further that the Komar mass as defined in \cite{Gauntlett:1998fz}:
\beqs
M_{K}&=&-\frac{3}{32\pi G}\int_{\infty} dS_{mn}D^{m}\bold{k}^n,
\label{KomarM}
\eeqs
is related to the counterterm mass $M$ in (\ref{masa}) by:
\beqs
2M_K=2M-{\cal T}L,
\eeqs
while the Komar angular momentum:
 \beqs
J_{K}&=&\frac{1}{16\pi G}\int_{\infty} dS_{mn}D^{m}\bold{m}^n,
\eeqs
is the same as the angular momentum computed using the counterterms method. Here $\bold{k}=\partial/\partial t$ is the canonically normalized timelike Killing vector, while $\bold{m}=\partial/\partial\psi$. 

As an example, let us consider first the case of a squashed black hole in the KK monopole background. Setting $a=0$, $j=0$ in the above formulae one obtains straightforwardly the conserved mass, charge and tension:
\beqs
M&=&\frac{\pi(r_0^4+2mr_0^2-3q^2)}{4\sqrt{r_0^4-2mr_0^2+q^2}},~~~~~Q=-\frac{\sqrt{3}\pi q}{2},~~~~~{\cal T}=\frac{r_0(r_0^2-m)}{4\sqrt{r_0^4-2mr_0^2+q^2}}.
\eeqs
Noting that the length of the $S^1$ circle at infinity is $L=2\pi r_0$ one can check directly that the Smarr relation is indeed satisfied:
\beqs
2M&=&3TS+\Phi Q+{\cal T}L,
\eeqs
where:
\beqs
\Phi&=&-\frac{\sqrt{3}q}{2}\frac{\sqrt{r_0^4-2mr_0^2+q^2}}{mr_0^2-q^2+r_0^2\sqrt{m^2-q^2}}
\eeqs
is the electric potential. Moreover, the first law of thermodynamics for the squashed black holes takes the simple form (\ref{firstlaw}). One should note here that the Komar mass as computed from (\ref{KomarM}) satisfies indeed the relation $2M_K=2M-{\cal T}L$, as expected from the generalized Smarr relation.

Take now the case of the rotating KK-monopole discussed in the previous section. Using the counterterms method one computes the mass, angular momentum and tension:\beqs
M&=&\frac{\pi r_0^2}{4G\sqrt{1-4j^2r_0^2}},~~~~~~~J_{\psi}=-\frac{\pi j^3r_0^6}{G},~~~~~~~{\cal T}=\frac{r_0(1-6j^2r_0^2)}{4G(1-4j^2r_0^2)},
\eeqs
while the dipole charge is given in (\ref{dipolecharge}). The entropy vanishes as expected since there is no event horizon in this geometry. The dipole potential can be read from the expected generalized Smarr relation:
\beqs
\Phi_D&=&\frac{2\sqrt{3}\pi jr_0^2}{G\sqrt{1-4j^2r_0^2}},
\eeqs
and it is an easy matter to check that the first law is indeed satisfied.

For the squashed Schwarzschild-G\"odel spacetime the conserved charges are given respectively by:
\beqs
M&=&\frac{\pi r_0}{4G}\frac{2m+r_0^2-32m^2j^2-24mr_0^2j^2+128m^3j^4+128m^2r_0^2j^4}{\sqrt{r_0^2-2m+16m^2j^2}\sqrt{1-8mj^2-4j^2r_0^2}},~~~D=\frac{\sqrt{3}}{2}jr_0^2,\nonumber\\{\cal T}&=&-\frac{m-r_0^2-16m^2j^2+6r_0^4j^2-64m^3j^4+64m^2r_0^2j^4}{4G\sqrt{r_0^2-2m+16m^2j^2}(1-8mj^2-4j^2r_0^2)},~~~~~~~J_{\psi}=-\frac{\pi j^3r_0^6}{G}.
\eeqs
The black hole horizon is located at:
\beqs
\rho_+&=&\frac{m(1-8mj^2)}{\sqrt{r_0^2-2m+16j^2m^2}},
\eeqs
and it has the topology of a squashed three-sphere. Its area can be easily evaluated and the Bekenstein-Hawking entropy of the black hole can be written as:
\beqs
S&=&\frac{4\pi^2}{G}\rho_+^2f(\rho_+)\sqrt{h(\rho_+)}=\frac{\pi^2}{G}\frac{r_0^2\sqrt{2m^3(1-8mj^2)^5}}{r_0^2-2m(1-8mj^2)}.
\eeqs
The angular velocity of the horizon becomes:
\beqs
\Omega&=&\frac{16r_0j^3\sqrt{r_0^2-2m+16m^2j^2}}{(1-8mj^2)^2\sqrt{1-8mj^2-4j^2r_0^2}}.
\eeqs
It is also straightforward to compute the Hawking temperature using the general formula in (\ref{temp}):
\beqs
T&=&\frac{\sqrt{r_0^2-2m(1-8mj^2)}\sqrt{1-4j^2r_0^2-8mj^2}}{2\pi r_0\sqrt{2m(1-8mj^2)^3}}.
\eeqs

With the dipole potential:
\beqs
\Phi_D&=&\frac{2\sqrt{3} \pi}{G}\frac{jr_0(1+16j^2r_0^4-16mj^2+64m^2j^4)}{(1-8mj^2)^2}\frac{\sqrt{r_0^2-2m+16m^2j^2}}{\sqrt{1-8mj^2-4j^2r_0^2}}
\eeqs
one can check that the generalized Smarr relation and the first law are indeed satisfied.

One of the main motivations of our work was studying the effect of the squashing factor on the properties of a black hole immersed in the G\"odel universe. Recall that a G\"odel black hole cannot have an arbitrarily large entropy as the area of the black hole horizon reaches a maximum value for some value of the mass parameter $m$ and it decreases when $m$ is further increased beyond that value \cite{Klemm:2004wq}. This is reminiscent of the case of a black hole in de Sitter universe, since its entropy is also bounded from above by the entropy of a Nariai black hole, which is the largest black hole that can fit within the cosmological horizon.

 One might expect that these intriguing properties of the black hole entropy in a pure G\"odel universe are lost once one introduces the squashing factor in the G\"odel geometry. However, it turns out that there still exist ranges of parameters such that this strange behavior is in fact preserved even in the squashed geometry.
\begin{figure}
\begin{center}
\includegraphics[angle=-90,totalheight=0.40\textheight,clip]{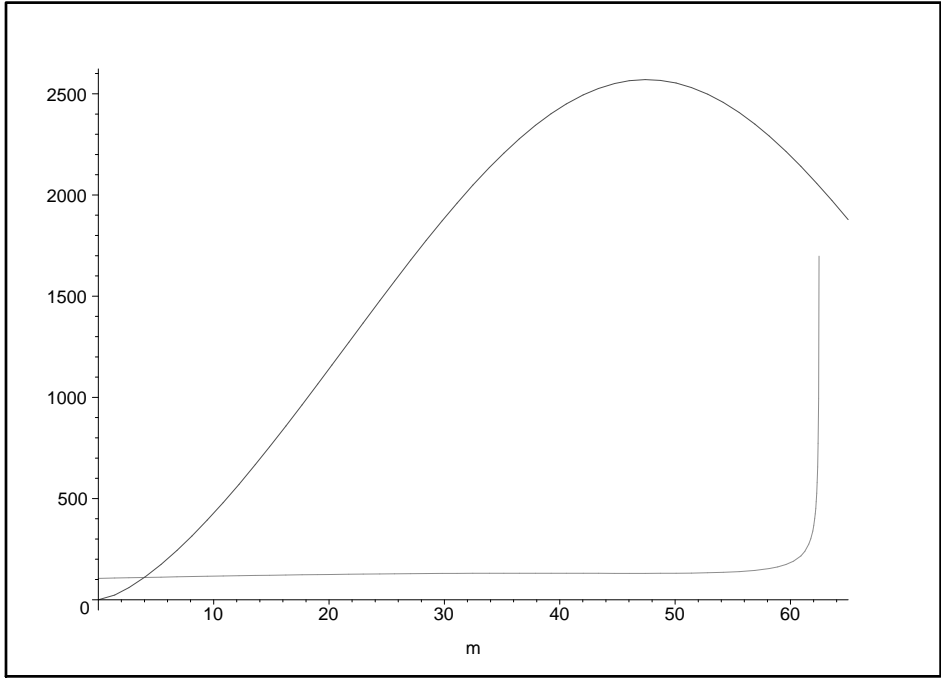}
\caption{Entropy $S$ and mass $M$ of a black hole in a squashed G\"odel universe for $r_0=10$ and $j=1/30$.}
\end{center}
\label{figure}
\end{figure}
For example, in Figure $1$ one notices that the entropy of a black hole in the squashed G\"odel background is indeed bounded from above and that there is a range of the mass parameter $m$ at fixed $j$ and $r_0$ for which the entropy decreases while the  conserved mass $M$ increases. A through investigation of the thermodynamic properties of these squashed black holes is outside the scope of this paper and we shall leave this interesting subject for future work.

\section{Conclusions}
 
 In the present paper we have considered the effect of the squashing transformation on the geometry of a general charged, rotating black hole embedded into the G\"odel universe.

We have seen that  the squashing function $\chi(r)$ is generally given in terms of a real parameter $c$, whose values control the squashing of the three-sphere at infinity. For negative values of $c=-r_0^2<0$ one recovers the solution given in the appendix of \cite{Tomizawa:2008hw}. By an appropriate change of coordinates it turns out that the region near $r_0$ corresponds in fact to an asymptotic region, however, we found that this region can be reached from both sides of $r_0$. The region $r<r_0$ corresponds to the so-called `left geometry' while the region $r>r_0$ gives a different background geometry. The `left' geometry asymptotes to a fully regular rotating Kaluza-Klein monopole background, while the `right' geometry corresponds to a similar rotating Kaluza-Klein monopole, however, with negative value of the NUT charge and it is therefore nakedly singular. By carefully choosing the parameters, one can ensure that either the left geometry or the right geometry become free of the closed timelike curves that plagued the initial un-squashed G\"odel geometry but not both of them at the same time.

We have considered the lift of the squashed geometry in ten dimensions and its $T$-dual geometry. It turns out that only in the case of unsquashed G\"odel itself does the ten-dimensional geometry factorize nicely into a $pp$-wave form. Interestingly, the lift of the general squashed rotating charged black hole in  G\"odel spacetime has as its $T$-dual geometry a rotating Kaluza-Klein black hole - like spacetime.

Using the counterterms method we computed the conserved charges and action of the squashed black hole. In particular, we found that the gravitational tension (per unit time) ${\cal T}$ can be naturally defined in this context as being the conserved quantity corresponding to the Killing vector $\partial/\partial\psi$. We have also considered the Komar quantities in terms of which the Smarr relation takes a particularly simple form, which, however, does not include the term involving the gravitational tension. To this end, we uncovered a simple relation between the conserved mass computed in the counterterms approach and the Komar mass: they are related by $2M_K-2M=-{\cal T}L$, where $L$ is the asymptotic length of the $\psi$ circle. If we use this relation in the expected generalized Smarr relation (\ref{smarr}), one finds that in the new Smarr relation the terms involving the gravitational tension cancel out.

Finally, we found that some unexpected thermodynamical properties of the uncharged black holes embedded in the G\"odel spacetimes can be preserved even after one introduces the squashing transformation. In particular, there are ranges of the parameters describing the black hole solution for which the black hole entropy is bounded from above and it can further decrease as one increases the mass of the black hole. We leave a more detailed analysis of the thermodynamical aspects of the general charged and rotating squashed black hole to future work.

\vspace{10pt}

{\Large Acknowledgements}

This work was supported by the Natural Sciences and Engineering Council of Canada.

\end{document}